# Development of a thorium coating on an aluminium substrate by using electrodeposition method and alpha spectroscopy

Dal-Ho Moon[1], Vivek Chavan[1], Vasant Bhoraskar[1], Yeong Hoon Jeong[1], Jung Ho Park[2], Su-Jeong Suh[2], and Seung-Woo Hong[1,3,*]

[1]*Department of Physics, Sungkyunkwan University, Suwon 16419, Rep. of Korea*
[2]*School of Advanced Materials Science and Engineering, Sungkyunkwan University, Suwon 16419, Rep. of Korea*
[3]*Rare Isotope Science Project, Institute for Basic Science, Daejeon 34000, Rep. of Korea*

A thin coating of thorium on aluminium substrates with the areal density of 110 to 130 μg/cm$^2$ is developed over a circular area of 22 mm diameter by using the electrodeposition method. An electrodeposition system is fabricated to consist of three components; an anode made of a platinum mesh, a cylindrical-shape vessel to contain the thorium solution, and a cathode in the form of a circular aluminium plate. The aluminium plate is mounted horizontally, and the platinum mesh is connected to an axial rod of an electric motor, mounted vertically and normal to the plane of the aluminium. The electrolyte solution is prepared by dissolving a known-weight thorium nitrate powder in 0.8 M HNO$_3$ and isopropanol. The system is operated either in constant voltage (CV) or constant current (CC) mode. Under the electric field between the anode and cathode, thorium ions were deposited on the aluminium substrate mounted on the cathode. In the CV mode at 320, 360, and 400 V and in the CC mode at 15 mA, thorium films were formed over a circular area of the aluminium substrate. The areal density of thorium coating was measured by detecting emitted alpha particles. The areal density of thorium varied from 80 to 130 μg/cm$^2$ by changing the deposition time from 10 to 60 min. The results from the CV mode and CC mode are compared, and the radial dependence in the measured areal density is discussed for different modes of the electric field. The developed thorium coatings are to be used in the in-house development of particle detectors, fast neutron converters, targets for thorium fission experiments, and other purposes.

*Corresponding author.
Email: swhong@skku.ac.kr

# 1. Introduction

A Time-of-Flight (TOF) facility for fast neutrons, called the Nuclear Data Production System (NDPS), capable of generating fast neutrons up to ~ 80 MeV, is constructed and under commissioning as part of the Rare Isotope Science Project (RISP) [1-4]. By using both white neutrons and quasi-monoenergetic neutrons, neutron-induced nuclear data for a variety of targets can be generated. To get accurate nuclear data, neutron detectors that can precisely measure and monitor the neutron flux are required.

The neutron detectors are normally based on the detection of secondary charged particles or fission fragments emitted from a material, called a neutron converter, through the reaction induced by the incident neutrons [5]. In such detectors, materials with a large fission cross section can be a good neutron converter. Uranium is often used as a neutron converter. However, uranium (even $^{238}$U) has the disadvantage of large fission cross sections for low energy neutrons as compared to thorium. Thus, as a fast neutron converter thorium can be preferred over uranium in such systems where a high flux of low energy neutrons exists as a background. Thorium has fission cross sections of ~mb or higher for neutrons of energies higher than 1 MeV, and it has in general a better social acceptance than uranium. Moreover, Th(n,f) cross sections are well-known for neutrons of high energies, and therefore thorium can be a good converter for a high energy neutron monitoring system. A good neutron converter needs to be have a well-characterised deposition of the converter material, such as the amount of deposition and uniformity [6]. Thus, thorium converters with a uniform areal density is required for the in-house development of neutron monitoring systems, which can be based on gaseous detectors such as the Parallel Plate Avalanche Counter (PPAC) [7] and MICRO-MEsh GAseous Structure (MICROMEGAS) detectors [8, 9]. The gaseous detectors such as PPAC and MICROMEGAS have the advantage of being able to measure neutron flux in real-time [10].

Thorium is a fertile material, which can breed a fissile isotope of $^{233}$U. Thus, many studies have been done for thorium fuel cycles [11, 12] and physical and chemical properties of thorium materials [13]. For example, the total kinetic energy and mass distribution of fission fragments of thorium were investigated to understand fission of thorium, induced by fast neutrons or high energy protons [14-16]. For such studies, high-quality thorium targets are required with known elemental concentration and uniformity [17]. There are studies of the electrodeposition method for making thorium targets with good uniformity of areal densities [18, 19]. The deposited thorium coatings can be used as a target for thorium fission experiments, fast neutron converters, production of radioisotopes, etc [20]. However, making targets of nuclear fuel materials such as thorium is rather a difficult task due to the constraints related to handling nuclear fuel material, safety measures, uniformity, and cost [21].

In the present work, an electrodeposition system was designed, fabricated, and used for making thorium coating on aluminium substrates for use as a thorium neutron converter or a thorium target for fission experiment. Studies of coatings of thorium on aluminium substrates by using the electrodeposition method were reported in the literature [19, 22-24], but there is no report on the areal density and homogeneity of the thorium coatings obtained under different conditions such as the constant voltage (CV) and constant current (CC) modes. We made thorium coatings on thin aluminium substrates by using CV and CC modes, and the results from different modes are compared to find an optimal condition.

## 2. Experiment
### 2.1 Electrodeposition system for thorium coatings

Figure 1 shows a graphical image and a photo of the electrodeposition system, designed and fabricated in the lab. It consists of three main parts; (i) an aluminium cathode fixed on an insulating stand (ii) a polyacetal cylinder with flanges on both ends mounted on the cathode stand, and (iii) an anode in the form of a platinum mesh, attached to the rod of a motor mounted on the top of the polyacetal cylinder. Polyacetal was used due to its high chemical resistance. The platinum mesh was connected electrically to the positive terminal of the power supply. At the bottom side of the polyacetal cylinder, a circular aluminium plate of 0.2 mm thick and 36 mm diameter was fixed on the cathode electrode and connected to the negative terminal of the power supply. Both the substrate and the cathode were made of aluminium. The distance between the anode and cathode was kept at 11 mm.

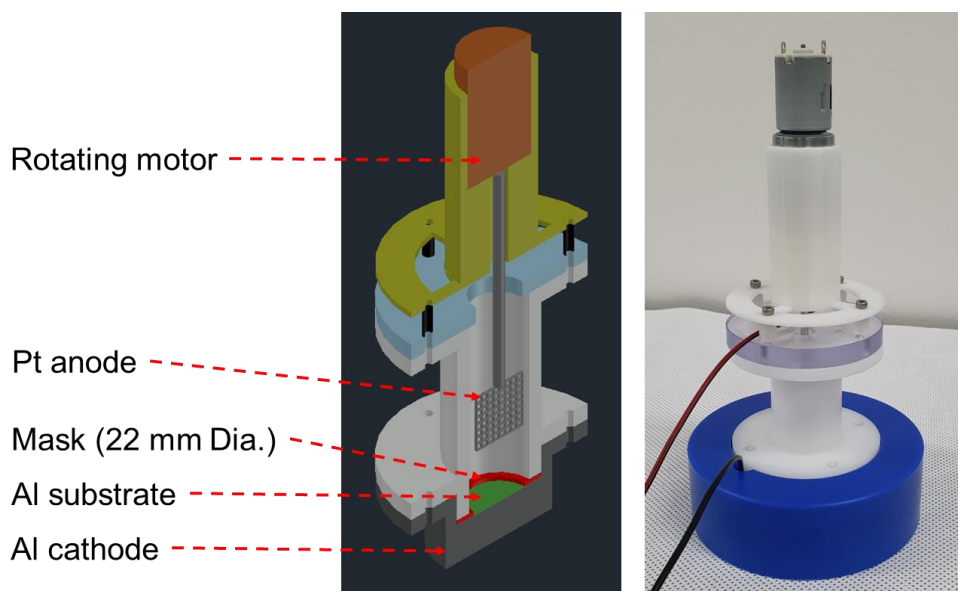

Fig. 1 The electrodeposition system for thorium coating on an aluminium substrate. The platinum mesh anode is connected to a rotating motor while the aluminium substrate and cathode are assembled at the bottom. The central cylindrical vessel of the system is filled with a thorium solution.

Thorium solution was made by following the process described in Refs. [17, 18, 25]. 5 mg of thorium nitrate powder was first dissolved in 30 μL $HNO_3$ of 0.8 M strength and then was dried at a temperature close to 200 °C for 30 min. The dried sample powder was dissolved in a mixture of 50 μL $HNO_3$ of 0.8 M strength and 12 ml of isopropanol. A polymer mask of 22 mm diameter was fixed on the top surface of the aluminium substrate. The central cylindrical-shape vessel of the system was filled with this thorium solution so that the column of the solution lies between the aluminum substrate on the bottom and the platinum mesh connected to the rotating motor. The thorium solution was stirred by rotating the platinum mesh at 10 rpm.

By applying a voltage to the respective electrodes, an electric field was formed between the anode and cathode. The thorium ions were deposited on the aluminium substrate and the area of the coating was controlled by a polymer mask. In this manner, thorium coating of a circular area of 22 mm diameter was obtained on aluminium substrates under the constant voltage (CV) and constant current (CC) modes. During the experiment, hydrogen bubbles can be formed in the solution because of heat and may interfere deposition of thorium ions on the substrate [26]. To avoid this problem, the aluminium cathode was cooled with water. In the CV mode, thorium coatings were obtained by applying 320, 360, and 400 V, in which cases the maximum and minimum values of the current were ~25 and ~10 mA, respectively, as will be shown in Section 3. In the CC mode, a current of 15 mA, which is roughly the average of the above-mentioned maximum and minimum current was used. Thorium coatings were obtained for different deposition periods such as 10, 20, 30, 40, 50, and 60 min. Each thorium-deposited aluminium substrate was removed from the solution and dried under an infrared lamp for 10 min to evaporate the residual solution on the aluminium substrate [27]. The areal density of each thorium coating obtained for different deposition time was measured and compared, as will be shown in Section 3. During the experiments the chemical components of thorium nitrate powder could be changed by absorbing moisture, and therefore all the experiments were conducted by keeping low humidity in the room with a humidity controller.

## 2.2 Characterization of the thorium coatings by the alpha spectroscopy

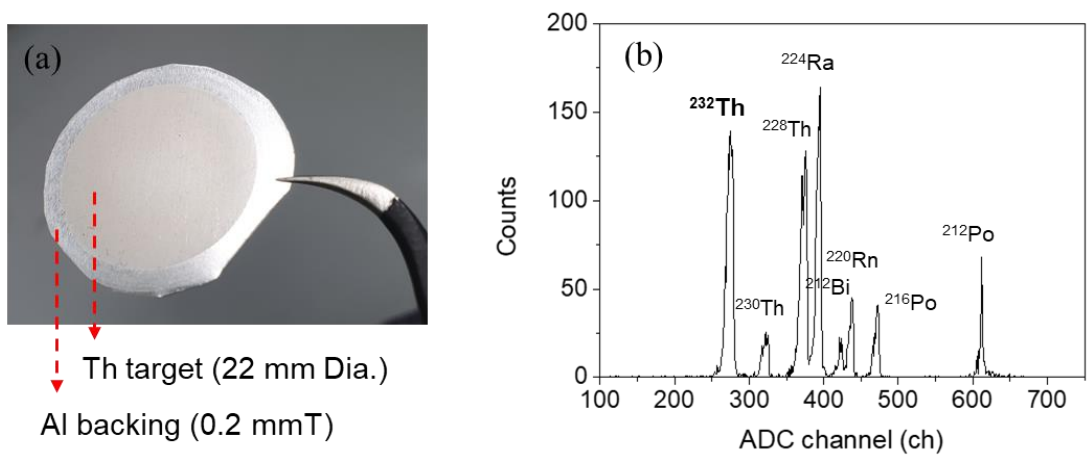

Fig. 2 (a) Thorium coating with 22 mm diameter deposited on a 0.2 mm thick Al substrate. (b) The alpha spectrum from the deposited thorium coating was recorded by a silicon detector.

A photo of a typical thorium film coated on an aluminium substrate is shown in Fig. 2(a). The alpha spectrometry was used for measuring the amount of deposited thorium on the aluminium substrate. The half-life of $^{232}$Th is ~ $1.4 \times 10^{10}$ years and it decays by emitting alpha particles of 4 MeV. An alpha spectrometer was assembled in the laboratory for measuring the alpha particle activity of the thorium coatings. It consists of a Passivated Implanted Planar Silicon (PIPS) detector mounted in a vacuum chamber and a sample holder. A thorium coated aluminium substrate was fixed in front of the silicon detector and the chamber was evacuated to a pressure of ~ $10^{-3}$ torr. The distance between the coated thorium and the silicon detector was 43 mm. The silicon detector was connected to a multichannel analyzer through a pre-amplifier and other electronic modules. A typical spectrum of the alpha particles emitted by the coated thorium is shown in Fig. 2(b). The first peak in the ADC channel is due to the decay alpha of $^{232}$Th, and other peaks are alpha events from other Th isotopes and daughter nuclides of $^{232}$Th [28]. The areal density ($\rho_A$) of the deposited thorium could be obtained by using the following expression [29],

$$\rho_A = \frac{M}{N_A \lambda} \frac{C}{A t \Omega_{eff}}, \tag{1}$$

where M is the molar mass of $^{232}$Th, $N_A$ is the Avogadro number, $\lambda$ is the decay constant of $^{232}$Th, C is the number of detected alphas in the first peak of Fig. 2(b), A is the deposition area, t is the measured time, and $\Omega_{eff}$ is the effective solid angle obtained by the GEANT4 simulation. In the simulation, we generated alpha particles in random directions from the thorium film and the effective solid angle was defined and calculated as the division of the number of events entering the silicon detector geometry by the total number of alpha particles generated. The effective solid angle subtended by the silicon detector was 0.048 $\pi$ sr.

The homogeneity of the thorium deposited on aluminium substrates for different deposition modes was measured by alpha spectrometry. Each sample was covered by a mask with a hole of 4 mm diameter and the spectrum of the alpha particles which passed through the mask hole was recorded. The alpha spectrum was recorded by shifting the position of the mask hole to three locations; center, 4.5 mm, and 9 mm away from the center in the radial direction.

## 3. Results and discussion

Figure 3(a) shows variations in the current between the anode and cathode in the CV mode at 320, 360, and 400 V, while Fig. 3(b) shows variations in the voltage between the anode and cathode in the CC mode, when the current was maintained at 15 mA.

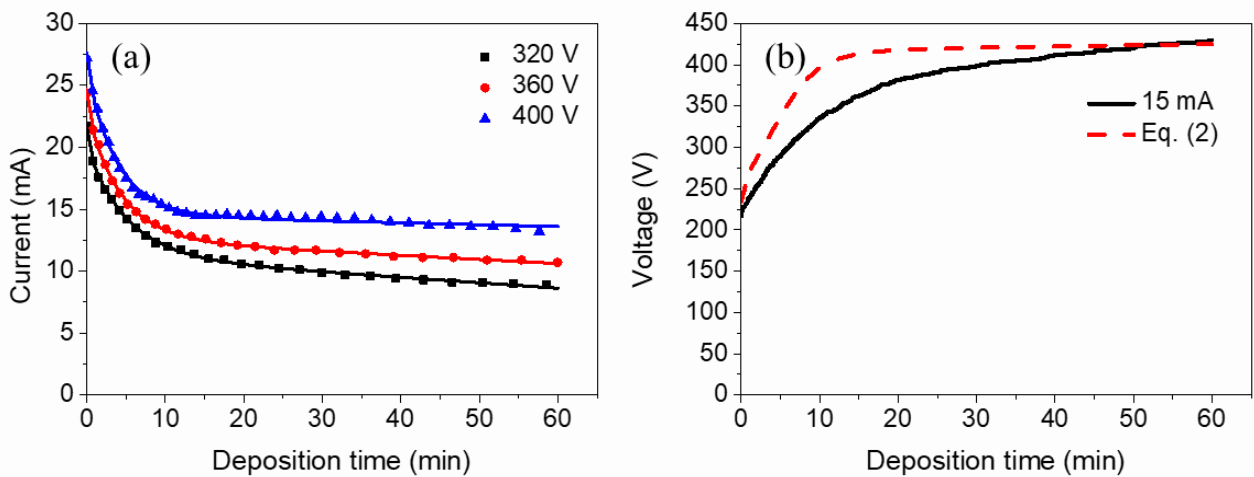

Fig. 3 (a) Variations in the current with the deposition time for 320, 360, and 400 V in the CV mode are plotted together with the fitted lines, and (b) variations in the voltage with deposition time in the CC mode are plotted by the solid curve and are compared with the dashed curve representing the voltages calculated by Eq. (2) with the parameters given in Table 1.

During the electrodeposition, the amount of thorium in the solution decreases with time, and therefore the electric conductivity of the solution also decreases. In addition, as the thickness of the thorium coating increases, the electric resistance due to the thorium coating increases. (The electric resistivity of thorium is larger than that of aluminium.) As seen in Fig. 3(a), initially the current decreases rapidly up to 20 min, and then decreases slowly. The rate of thorium deposition is, therefore, expected to be high up to about 20 min, and then decreases with the deposition time. A similar behavior can be seen for the CC mode. As shown in Fig. 3(b), the voltage for the CC mode starts initially with around 225 V and rapidly increases up to 20 min deposition time because the electric conductivity of the solution is high in the early stage and the electric resistance increases as the thorium film builds up. The voltage increases slowly after 20 min and reaches 420 V at the 60 min deposition time. Both the current for the CV mode and the voltage for the CC mode become plateaus as the amount of deposited thorium reaches a saturation point.

The measured currents plotted by the symbols in Fig. 3(a) for the CV mode of 320, 360, and 400 V were fitted with the following expression:

$$I(V) = I_1(V)exp\left(-\frac{t}{\tau_1(V)}\right) + I_2 exp\left(-\frac{t}{\tau_2}\right) + I_3(V) + I_4(V)t, \qquad (2)$$

where I(V) is the current according to a bias voltage $V$, $t$ is the deposition time, and $I_i(i = 1\sim4)$, $\tau_j(j = 1,2)$ are the adjustable parameters in the exponential and linear functions. The parameters in Eq. (2) can be determined for each voltage of 320, 360, and 400 V and the values are summarized in Table 1 together with the resulting chi-square values. The curves in Fig. 3(a) show that the currents decreased exponentially at the beginning of the deposition. The fact that $I_2 = 1.6$ mA while $I_1(V) \approx 10$ mA shows that the general behavior of the curves in Fig. 3(a) could be determined by the first exponential term. However, to fit the measured curves well enough to get small values of chi-squares, additional terms are needed. After ~20 min, exponential terms become negligible, and the currents decrease linearly. The decrease in the current implies an increase in the resistance and in the amount of the deposited thorium. Thus, based on the current variation, most of the thorium is expected to be deposited in the early stage.

Table 1 The fitting parameters in Eq. (2)

| CV mode | $I_1(V)$ (mA) | $I_2$ (mA) | $I_3(V)$ (mA) | $I_4(V)$ (mA·min) | $\tau_1(V)$ (min) | $\tau_2$ (min) | $\chi^2$ |
|---|---|---|---|---|---|---|---|
| 320 V | 8.7 | 1.6 | 11.2 | -0.04351 | 5.2 | 0.32 | 0.01 |
| 360 V | 10.2 | 1.6 | 12.6 | -0.03338 | 4.3 | 0.32 | 0.01 |
| 400 V | 11.5 | 1.6 | 14.5 | -0.01484 | 4.0 | 0.32 | 0.05 |

By using Eq. (2) with the parameters in Table 1, the changes in the voltage for the CC mode as a function of the deposition time can be calculated for a fixed value of I(V) = 15 mA. In the calculations, the parameters summarized in Table 1 were assumed to have a linear relationship with respect to the voltage. The calculated voltages in the CC mode are plotted by the dashed curve in Fig. 3(b) and compared with the measured voltage plotted by the solid curve. The calculated voltage and the measured voltage agree initially, but the voltage calculated by Eq. (2) obtained for the CV modes becomes larger than the voltage measured for the CC mode, as the deposition time increases. On the other hand, as the deposition time further increases and reaches 60 min, both the voltage calculated from Eq. (2) and that measured for the CC mode converge to the same value. When the deposition starts, the amount of deposited thorium is small and the physical and chemical conditions of the thorium solution and the cathode are about the same. Thus, the voltages and currents through the thorium film do not depend on the mode. However, as the thorium film on the cathode grows, the voltages and currents from the CV mode is not the same as those from the CC mode, due to the difference in the physical and chemical conditions. When the thickness of the thorium film reaches a saturation, again the physical and chemical conditions are about the same, and the voltages and currents do not depend on the (CV or CC) mode.

For a high-quality target production, it is important to have a uniform areal density. When the electric field is uniformly formed between the electrodes, the current density is also formed evenly on the Al substrate and a homogeneous thorium coating can be obtained. If the electric field is not uniform, thorium is not expected to be well deposited in the area where the current density is low, which may cause low areal density of thorium. Thus, it is necessary to know the optimized deposition condition for this electrodeposition setup. The areal density and homogeneity of deposition were measured by detecting alphas emitted from the thorium, and the results were compared for each deposition mode.

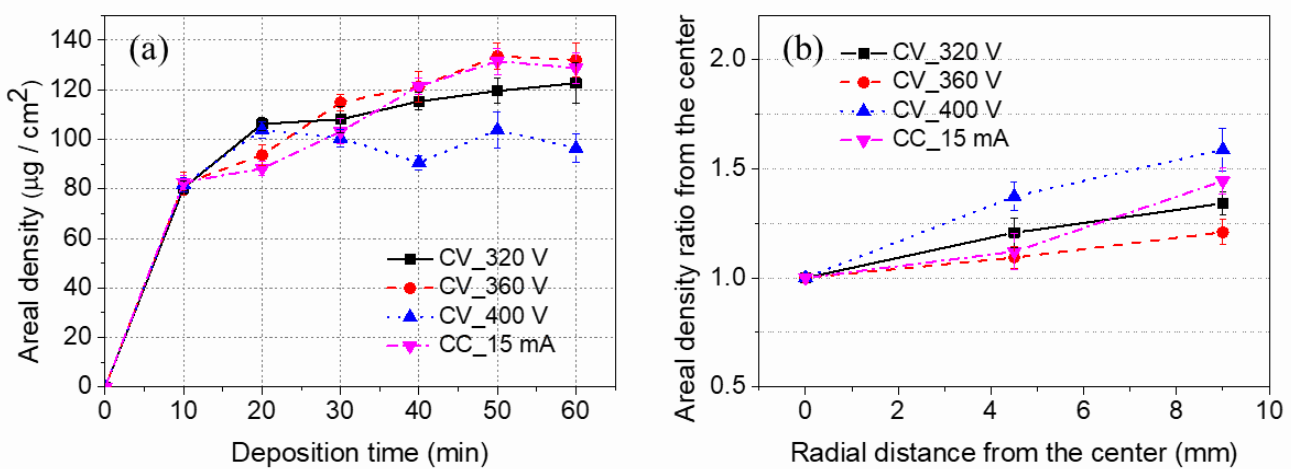

Fig. 4 (a) Variation of the areal density of the deposited thorium with deposition time and (b) homogeneity of coating measured as the ratio of the areal density at different radial distances to that at the center. Only the statistical errors of alpha spectrometry are included in the error bar.

Figure 4(a) shows the areal density of the deposited thorium measured by the alpha spectrometry. The areal density reached around 80 μg/cm$^2$ in 10 min of the deposition time at 320, 360, and 400 V in the CV mode as well as in the CC mode. This shows that the initial growth of the thorium film does not depend much on the conditions such as the applied voltages and currents, when the thorium film is thin enough. This is consistent with the results shown in Fig. 3. The areal density for all modes increases rapidly until ~20 min except for the case of 400 V in the CV mode, and then slowly increases to 120 ~ 130 μg/cm$^2$ at t = 60 min. Such a behavior of saturation in the areal density is similar to the saturation behavior of the voltage in the CC mode as shown in Fig. 3(b). In the CV mode of 400 V, the areal density reached close to 100 μg/cm$^2$ in 20 min of the deposition time and then saturated with some fluctuations during the rest of the deposition time up to 60 min. We may speculate that under the strong electric field produced by 400 V, the thorium ions moved with larger kinetic energies and therefore the thorium coating already deposited on the aluminium substrate was bombarded with the following thorium ions of large energies. In this process, a large number of secondary electrons could have been produced from the thorium coating, which neutralized a fraction of the incoming thorium ions. In addition, due to the impact of the energetic thorium ions, a fraction of the deposited thorium atoms on the substrate could have been removed from the coating due to the breaking of bonds. As a result, thorium coating could not further grow at 400 V after t ≈ 20 min.

Figure 4(b) shows that the ratio of the areal density at different radial distances to that at the center increased with the radial distance for all three voltages of 320, 360, and 400 V in the CV mode, as well as in the CC mode at 15 mA. In the current experimental setup, the most uniform coating was produced at 360 V in the CV mode with about 25 % difference in thickness between the center and the radial distance of 11 mm, whereas the inhomogeneity was highest for 400 V in the CV mode. In the CV mode of 400 V, it was observed that thorium was deposited more at the edge, and also the maximum areal density was low. In the case of 15 mA of the CC mode, the thorium uniformity is similar to that from the CV mode at 360 V at the radial distance of 4.5 mm. However, at a distance of 9 mm, the ratio becomes 1.4, which is much higher than the ratio of 1.2 obtained for 360 V in the CV mode.

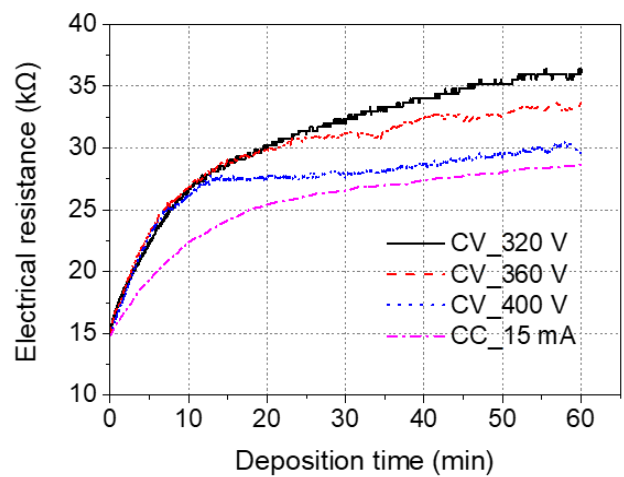

Fig. 5 The electrical resistance in each deposition mode as a function of the deposition time.

From Figs. 3(a) and 3(b), the electrical resistances can be obtained and plotted in Fig. 5. Regardless of the deposition mode, the values of the electrical resistances are the same as 15 kΩ at the moment when deposition starts as shown in Fig. 5. The resistances from the CV modes increase almost the same up to 10 min, but for 400 V the resistance changes only slightly after 10 min deposition. The fact that the resistance remains more or less the same for 400 V is consistent with the saturation behavior of the areal density for 400 V in Fig. 4(a) (Note that the areal density for the CV mode with 400 V is more or less constant after 10 min). In other deposition modes except for 400 V of the CV mode, the resistance increases still further after 10 min. The resistance was the highest in the case of 320 V of the CV mode, and was the lowest for the CC mode with 15 mA. The areal density obtained from the alpha spectroscopy for the CC mode is similar to that for the CV mode with 360 V. However, the resistance for the CC mode is much smaller than that for the CV mode with 360 V. The lowest electrical resistance for the CC mode may be due to the lowest porosity of the thorium sample [30] compared with other deposition modes.

In the past, thorium targets were prepared by using similar electrodeposition method [19, 22-24]. However, there was no study comparing the areal density, homogeneity and electrical properties of the system by changing the deposition conditions, such as the CV and the CC modes. In our electrodeposition system, we found that the condition of 360 V in the CV mode is an optimized setup by considering the thickness and homogeneity of the deposited thorium. The coating quality produced by using electrodeposition depends on the electric field, deposition time, etc. Thus, it seems necessary to find an optimal condition for each specific electrodeposition system.

## 4. Conclusion

An electrodeposition system was developed for coating thorium on aluminium substrates by using the CV and the CC modes. The electrodeposition condition was optimized by trying different voltages, 320, 360, and 400 V in the CV mode and 15 mA current in the CC mode. The uniformity of the thorium coating was studied by alpha spectroscopy. The highest areal density and the maximum homogeneity was obtained when the deposition was carried out at 360 V in the CV mode with our system. The optimized electrodeposition conditions can be used to prepare thorium samples which can be used as a neutron convertor for the fast neutrons to monitor the neutron beam profile of TOF facilities as well as for other applications where thin thorium coatings are required.

**Acknowledgments**

This work was supported in part by the National Research Foundation of Korea (NRF) funded by the Ministry of Science and ICT (Grant nos. NRF-2013M7A1A1075764, NRF-2015H1D3A1066285, NRF-2016R1D1A1B03935429, NRF-2018M2A8A2083829, and NRF-2019R1I1A1A01060797).

# References

1. B. H. Kang et al., *ISOL facility for rare isotope beams at RAON,* Journal of the Korean Physical Society **63**, (2013).
2. S. Jeong, P. Papakonstantinou, H. Ishiyama, and Y. Kim, *A Brief Overview of RAON Physics,* Journal of the Korean Physical Society **73**, 4 (2018).
3. Y. J. Kim, *Current status of experimental facilities at RAON,* Nuclear Instruments and Methods in Physics Research Section B: Beam Interactions with Materials and Atoms **463**, (2020).
4. Y. Chung, H. Kim, and M. Kwon, *Current status of the RAON low-energy heavy ion accelerator,* Journal of the Korean Physical Society **80**, 8 (2022).
5. S. Andriamonje et al., *Experimental studies of a Micromegas neutron detector,* Nuclear Instruments and Methods in Physics Research Section A: Accelerators, Spectrometers, Detectors and Associated Equipment **481**, 1-3 (2002).
6. R. Beyer et al., *Characterization of the neutron beam at nELBE,* Nuclear Instruments and Methods in Physics Research Section A: Accelerators, Spectrometers, Detectors and Associated Equipment **723**, (2013).
7. C. Akers et al., *Large area PPAC development at the rare isotope science project,* Nuclear Instruments and Methods in Physics Research Section A: Accelerators, Spectrometers, Detectors and Associated Equipment **910**, (2018).
8. D. Y. Kim et al., *Development of the MICROMEGAS detector for measuring the energy spectrum of alpha particles by using a 241 Am source,* Journal of the Korean Physical Society **68**, (2016).
9. C. Ham et al., *A Simulation Study and Its Experimental Validation for the Detection of Neutrons with a Continuous Energy Spectrum by Using a MICROMEGAS Detector,* Journal of the Korean Physical Society **75**, (2019).
10. S. Andriamonje et al., *Development and performance of Microbulk Micromegas detectors,* Journal of Instrumentation **5**, 02 (2010).
11. P. Rodriguez and C. V. Sundaram, *Nuclear and materials aspects of the thorium fuel cycle,* Journal of Nuclear Materials **100**, 1 (1981).
12. D. Heuer et al., *Towards the thorium fuel cycle with molten salt fast reactors,* Annals of Nuclear Energy **64**, (2014).
13. U. E. Humphrey and M. U. Khandaker, *Viability of thorium-based nuclear fuel cycle for the next generation nuclear reactor: Issues and prospects,* Renewable and Sustainable Energy Reviews **97**, (2018).
14. M. C. Duijvestijn et al., *Proton-induced fission at 190 MeV of nat W, 197 Au, nat Pb, 208 Pb, and 232 Th,* Physical Review C **59**, 2 (1999).
15. I. V. Ryzhov et al., *Fragment-mass distributions in neutron-induced fission of Th 232 and U 238 at 33, 45, and 60 MeV,* Physical Review C **83**, 5 (2011).
16. D. Tarrio, *Neutron-induced fission fragment angular distribution at CERN n TOF: The Th-232 case.* 2012, Ph.D. Thesis, University of Santiago de Compostela.
17. G. Sibbens et al., in *AIP Conference Proceedings*. (2018).
18. C. Ingelbrecht, A. Moens, R. Eykens, and A. Dean, *Improved electrodeposited actinide layers,* Nuclear Instruments and Methods in Physics Research Section A: Accelerators, Spectrometers, Detectors and Associated Equipment **397**, 1 (1997).
19. Y. He et al., *Molecular plating of actinide compounds on wafer-scale aluminum substrate,* Journal of Alloys and Compounds **878**, (2021).
20. G. Sibbens et al., *Preparation of 240Pu and 242Pu targets to improve cross-section measurements for advanced reactors and fuel cycles,* Journal of Radioanalytical and Nuclear Chemistry **299**, 2 (2014).
21. G. Sibbens, A. Moens, D. Vanleeuw, D. Lewis, and Y. Aregbe, in *EPJ Web of Conferences*. EDP Sciences (2017).
22. D. Roman, *The electrodeposition of thorium in natural materials for alpha spectrometry,* Journal of Radioanalytical and Nuclear Chemistry **60**, 2 (1980).
23. L. Hallstadius, *A method for the electrodeposition of actinides,* Nuclear Instruments and Methods in Physics Research **223**, 2-3 (1984).
24. L. C. Hao et al., *Rapid preparation of Uranium and Thorium alpha sources by electroplating technique,* Kerntechnik **75**, 6 (2010).

25. W. S. Aaron, M. Petek, L. A. Zevenbergen, and J. R. Gibson, *Development and preparation of thin, supported targets and stripper foils,* Nuclear Instruments and Methods in Physics Research Section A: Accelerators, Spectrometers, Detectors and Associated Equipment **282**, 1 (1989).
26. V. Jobbágy et al., *Preparation of high-resolution 238U α-sources by electrodeposition: a comprehensive study,* Journal of Radioanalytical and Nuclear Chemistry **298**, 1 (2013).
27. J. C. Harduin, B. Peleau, and D. Levavasseur, *Détermination analytique des actinides dans les milieux biologiques,* Radioprotection **31**, 2 (1996).
28. S. Srivastava et al., *Concentration and activity ratio of thorium isotopes in surface soil around proposed uranium mining site in India,* Radiation Protection and Environment **36**, 3 (2013).
29. J. Strišovská, J. Kuruc, D. Galanda, and Ľ. Mátel, *Surface's weights of electrodeposited thorium samples determined by alpha spectrometry,* Journal of Radioanalytical and Nuclear Chemistry **288**, 2 (2011).
30. M. Hakamada, T. Kuromura, Y. Chen, H. Kusuda, and M. Mabuchi, *Influence of porosity and pore size on electrical resistivity of porous aluminum produced by spacer method,* Materials transactions **48**, 1 (2007).